\documentclass[
preprintnumbers,
tightenlines
,showpacs
,showkeys
,nofootinbib
,aps
,amsfonts
,amssymb
]{revtex4}

\usepackage{amsmath}
\usepackage{graphicx}
\usepackage{epic}
\usepackage{eepic}
\usepackage{epsfig}
\usepackage{latexsym}
\usepackage{subfigure}
\usepackage{float}
\usepackage[all]{xy}
\usepackage{amsfonts}
\usepackage{amssymb} 
\usepackage{color}
\usepackage{multirow} 

\newcommand{\bce}{\begin{center}}  
\newcommand{\ece}{\end{center}}
\newcommand{\beq}{\begin{equation}}  
\newcommand{\eeq}{\end{equation}}
\newcommand{\beqy}{\begin{eqnarray}}
\newcommand{\eeqy}{\end{eqnarray}}

\def\){\right)} 
\def\({\left(} 
\def\]{\right]} 
\def\[{\left[}

\begin{document}
\preprint{IMSC/2009/12/14}
\title{Chiral and Diquark condensates at large magnetic field\\
 in two-flavor superconducting quark matter}
\author{Tanumoy Mandal$^1$, Prashanth Jaikumar$^{1,2}$, Sanatan Digal$^1$}
\affiliation{Institute of Mathematical Sciences, C.I.T Campus, Chennai, TN 600113, India}
\affiliation{California State University Long Beach, Long Beach, CA 90840 USA}

\begin{abstract} 
We study the effect of a large magnetic field on the chiral and diquark condensates in a regime of moderately dense quark matter. Our focus is on the inter-dependence of the two condensates through non-perturbative quark mass and strong coupling effects, which we address in a 2-flavor Nambu-Jona-Lasinio (NJL) model. For magnetic fields $eB\lesssim 0.01$ GeV$^2$ (corresponding to $B\lesssim 10^{18}$G), our results agree qualitatively with the zero-field study of Huang et al., who found a mixed broken phase region where the chiral and superconducting gap are both non-zero. For $eB\gtrsim 0.01$ GeV$^2$ and moderate diquark-to-scalar coupling ratio $G_D/G_S$, we find that the chiral and superconducting transitions become weaker but with little change in either transition density. For large $G_D/G_S$ however, such a large magnetic field disrupts the mixed broken phase region and changes a smooth crossover found in the zero-field case to a first-order transition at neutron star interior densities. 

\end{abstract}

\maketitle 
\section{Introduction}
\label{sec_intro}

The existence of deconfined quark matter in the dense interior of a neutron star is an interesting question that has spurred research in several new directions in nuclear astrophysics. On the theoretical side, it has been realized~\cite{Alford:1998mk,Bailin,Rapp:1997zu} that cold and dense
quark matter must be in a superfluid state with many possible
intervening phases~\cite{Steiner:2002gx,Schmitt:2004hg,Shovkovy:2004um,Sharma:2006ig,Neumann:2002jm} between a few times nuclear matter density to
asymptotically high density, where quarks and gluon interact weakly. 
The observational impact of these phases
on neutron star properties can be varied and dramatic~\cite{Alc,Page:2002bj,Glendenning:1997fy,Jaikumar:2002vg,Reddy:2002xc,Jaikumar:2008kh}. 
Therefore, it is of interest to situate theoretical ideas and advances 
in our understanding of dense quark matter in the context of neutron
stars, which serve as unique astrophysical laboratories for
such efforts. 

\vskip 0.2cm

The phenomenon of color superconductivity in Quantum Chromodynamics (QCD)
at large quark chemical potential $\mu$ is interesting for several reasons.
Weakly attractive gluonic forces drive a BCS instability~\cite{Bardeen}
which results in the formation of Cooper pairs of quarks, breaking 
several global symmetries spontaneously and resulting in the appearance
of (pseudo-)Nambu-Goldstone modes~\cite{Alford:1998mk,Rapp:1997zu}. 
Quarks and gluons can
acquire an energy gap whose magnitude depends on the baryon density,
pairing channel, quark masses, flavor of participating quarks, neutrality
constraints etc (see~\cite{Alford:2007xm} for a recent comprehensive review). 
New massless modes arise in this phase, such as the 
superfluid phonon (corresponding to gentle fluctuations of the baryon number) 
and a rotated photon (a linear combination of the vacuum $U(1)_{\rm em}$ 
photon and gluonic hypercharge). 
The collective response of such modes to external perturbations determines the
transport properties of the diquark phase, with many phenomenological
consequences for neutron stars. 

\vskip 0.2cm
 
At asymptotically high density $\mu\gg \Lambda_{\rm QCD}$ and for number of flavors $N_f$=3, 
the preferred pairing pattern is a flavor and color-democratic one termed the 
color-flavor-locked (CFL) phase~\cite{Alford:1998mk}. This idealized phase, while it displays the essentially novel features of the color superconducting state,
is unlikely to apply to the bulk of neutron star matter, since even ten times 
nuclear matter saturation density ($\rho_0$) only corresponds to a quark chemical potential $\mu\sim 500$ MeV$\gg\hskip -0.4cm/~\Lambda_{\rm QCD}$ (assuming 3 massless flavors). At these densities, quark mass effects can be important,
and as pointed out in~\cite{Huang:2001yw}, must be treated non-perturbatively. It is reasonable to think that the strange quark current mass, being much larger than that of the up and down quarks, inhibits pairing of strange quarks with light quarks. For the purpose of this work, we will therefore adopt the scenario of quark matter in the two-flavor superconducting phase, which breaks the color $SU(3)$ symmetry to color $SU(2)$, leaving light quarks of one color (say ``3'') and all colors of the strange quark unpaired. Although this phase appears to be disfavored in compact stars~\cite{Alford:2002kj, Aguilera:2004ag}, we adopt it here to highlight the competition between the chiral and diquark condensates in the most straightforward way, without the additional complications of compact star constraints. Also, our results will be qualitatively true for the 2SC+s phase~\cite{Steiner:2002gx,Mishra:2004gw}, which can be studied similarly by simply embedding the strange quark, which is inert with respect to pairing, in the enlarged 3-flavor space.   

\vskip 0.2cm

Our objective in this letter is a numerical study of the competition between chiral and diquark condensates at moderately large $\mu$ and large magnetic field using the NJL model, similar in some respects to previous works~\cite{Huang:2001yw,Schwarz:1999dj,Mishra:2004gw,Mishra:2003nr}, which treat the quark mass non-perturbatively. Instanton-based calculations and random-matrix methods have also been employed in studying the interplay of condensates~\cite{Berges:1998rc,Carter:1999xb,Vanderheyden:1999xp}. In essence, smearing of the Fermi surface by diquark pairing can affect the onset of chiral symmetry restoration, which happens at $\mu\sim M_q$,
where $M_q$ is the constituent quark mass scale~\cite{Chen:2008zr}. Since $M_q$ appears also in the (Nambu-Gorkov) quark propagators in the gap equations, a coupled analysis of chiral and diquark condensates is required. This was done for the 2-flavor case with a common chemical potential in~\cite{Huang:2001yw}, but for zero magnetic field. We use a self-consistent approach to calculating the condensates from the coupled gap equations, and find small quantitative (but not qualitative) differences from the results of Huang et al~\cite{Huang:2001yw} for zero magnetic field. This small difference is most likely attributed to a difference in numerical procedures in solving the gap equations. We also address the physics of chiral and diquark condensates affected by large in-medium magnetic fields that are generated by circulating currents in the core of a neutron or hybrid star. Magnetic fields as large as $10^{15}$G are expected on the surface of magnetars and their interior field may be as large as $10^{18}$G, pushing the limits of structural stability of the star~\cite{Bocquet:1995je,Broderick:2001qw}. There is no Meissner effect for the rotated photon, which has only a small gluonic component, therefore, magnetic flux is hardly screened~\cite{Alford:1999pb}, implying that studies of magnetic effects in color superconductivity are highly relevant~\cite{Ferrer:2005pu,Ferrer:2006vw}. Including the magnetic interaction of the quarks with the external field leads to qualitatively different features in the competition between the two condensates, and this is the main result of our work. 

\vskip 0.2cm  

In Section \ref{sec_lag}, we state the model NJL Lagrangian and recast the partition function in terms of interpolating bosonic variables. In Section \ref{sec_gap}, we obtain the gap equations for the chiral and diquark order parameters by minimizing the thermodynamic potential (we work at zero temperature throughout since $T_{\rm star}\ll \mu$). In Section \ref{nu_ana}, we discuss our numerical results for the coupled evolution of the condensates as a function of a single ratio of couplings, chemical potential and magnetic field before concluding in Section \ref{sec_conc}. 

\section{Lagrangian and Thermodynamics}
\label{sec_lag}

We employ a Lagrangian density for two quark flavors ($N_f$=2) applicable to the scalar-isoscalar and the pseudoscalar mesons and scalar diquarks

\beqy
&~&~~~~~~~~{\cal L}_{NJL}={\cal L}_{kin}+{\cal L}_{\bar{q}q}+{\cal L}_{qq}\,;~~\quad {\cal L}_{kin}=\bar{q}(i\hskip -0.1cm\not\!\partial+\mu\gamma^{0}-\hat{m})q\,,\nonumber\\
&~&{\cal L}_{\bar{q}q}=G_{S}[(\bar{q}q)^2+(\bar{q}i\gamma_{5}\vec{\tau}q)^2]\,,~~\quad
{\cal L}_{qq}=G_{D}(\bar{q}i\gamma_{5}\epsilon_{f}\epsilon_{c}q^C)(\bar{q}^{C}i\gamma_{5}\epsilon_{f}\epsilon_{c}q)\,,
\eeqy

where $q=(u,d)$, $\mu$ is the common quark chemical potential\footnote{Our assumption of a common chemical potential is for simplicity; in an actual neutron star containing some fraction of neutral 2SC or 2SC+s quark matter in beta-equilibrium, additional chemical potentials for electric charge and color hypercharges must be introduced in the NJL model. Furthermore there can be more than one diquark condensate and in general $M_u\neq M_d\neq M_s$~\cite{Mishra:2004gw}.}, $\hat{m}={\rm diag}(m_{u},m_{d})$ is the current quark mass matrix in the flavor basis, $\vec{\tau}=(\tau^1,\tau^2,\tau^3)$ are the Pauli matrices in flavor space, $(\epsilon_f)_{ij}$ and $(\epsilon_c)^{\alpha\beta 3}$ are antisymmetric matrices for flavor and color, while $\bar{q}^C=-q^TC$ and $q^C=C\bar{q}^T$ with charge-conjugation matrix $C=-i\gamma^0\gamma^2$. We take $m_{u}=m_{d}=m_{0}\neq 0$. As the diquark coupling $G_D$ and scalar coupling $G_S$ depend on the form of the underlying interaction and are not universal, we fix $G_S$ empirically and vary $G_D/G_S$ as a parameter, within phenomenological bounds~\cite{Berges:1998rc}. Introducing the bosonic fields $\sigma: (\bar{q}q), ~\vec{\pi}: (\bar{q}i\gamma_5\vec{\tau}q), ~\Delta: (\bar{q}^{C}i\gamma_{5}\epsilon_{f}\epsilon_{c}q), ~{\rm and}~\Delta^{*}:(\bar{q}i\gamma_{5}\epsilon_{f}\epsilon_{c}q^C)$, the bosonized Lagrangian becomes

\beq
{\cal L}_{b}=\bar{q}(i\hskip -0.1cm\not\!\partial+\mu\gamma^0-\hat{m})q-\bar{q}(\sigma+i\gamma_5\vec{\pi}\cdot\vec{\tau})q-\frac{1}{2}\Delta^{*}(\bar{q}^{C}i\gamma_{5}\epsilon_{f}\epsilon_{c}q)-\frac{1}{2}(\bar{q}i\gamma_{5}\epsilon_{f}\epsilon_{c}q^C)\Delta-\frac{\sigma^2+
\vec{\pi}^2}{4G_s}-\frac{\Delta^{*}\Delta}{4G_D}\,.
\eeq

Chiral symmetry breaking and color superconductivity in the 2SC phase is manifest by non-vanishing VEVs for $\sigma$ and $\Delta$ (we do not include the possibility of pion condensation~\cite{Andersen:2007qv}). The partition function in the presence of an external magnetic field $B$ is given by  

\beqy
\mathcal{Z}&=&N\int[d\bar{q}][dq]\textrm{exp}\Bigl\lbrace\int_{0}^{\beta}d\tau\int d^{3}\vec{x}(\mathcal{L}_{b}+\mathcal{L}_{em})\Bigr\rbrace~~~{\rm where}~~~\mathcal{L}_{em}=\frac{1}{2}\tilde{e}\tilde{Q}(\bar{q}\kern+0.30em /\kern-0.70em{A}q-\bar{q}^{C}\kern+0.30em /\kern-0.70em{A}q^{C})-\frac{B^{2}}{8\pi}\,,
\eeqy

which can be rewritten in terms of the bosonized version as

\beqy
\mathcal{Z}&=&\mathcal{Z}_{c}\mathcal{Z}_{1,2}\mathcal{Z}_{3}~~~{\rm where}~~~\mathcal{Z}_{c}=N~{\rm exp}\left\{-\int_0^{\beta}d\tau\int d^3x\left[\frac{\sigma^2}{4G_s}+\frac{\Delta^{2}}{4G_D}+\frac{B^{2}}{8\pi}\right]\right\}\,,\\\nonumber
\mathcal{Z}_{1,2}&=&\int[d\bar{Q}][dQ]{\rm exp}\left\{\int_0^{\beta}d\tau\int d^3x\left[\frac{1}{2}{\cal L}_{\rm kin}(Q,Q^{c})+\frac{1}{2}\tilde{e}\tilde{Q}(\bar{Q}\kern+0.30em /\kern-0.70em {A}Q-\bar{Q}^c\kern+0.30em /\kern-0.70em {A}Q^c)+\frac{1}{2}\bar{Q}\Delta^-Q^c+\frac{1}{2}\bar{Q}^c\Delta^+Q
\right]\right\}\,,\\ \nonumber
\mathcal{Z}_{3}&=&\int[d\bar{q}_3][dq_3]{\rm exp}\left\{\int_0^{\beta}d\tau\int d^3x\left[\frac{1}{2}{\cal L}_{\rm kin}(q_{3},q_{3}^{c})+\frac{1}{2}\tilde{e}\tilde{Q}(\bar{q}_3\kern+0.30em /\kern-0.70em{A}q_3-\bar{q}_3^c \kern+0.30em /\kern-0.70em{A}q_3^c)\right]\right\}\,.\\ \nonumber
\eeqy

The bosonized part $\mathcal{Z}_{c}$ serves as a constant multiplicative factor. The subscripts ``$1,2$'' refer to quarks of color 1 and 2 with $Q=q_{1,2}$ and $``3"$ to quarks of color 3. The kinetic operators in $\mathcal{L}_{kin}(q,q^{c})$ now read $(i\hskip -0.1cm\not\!\partial+\mu\gamma^0-M)$ where $M=m_0+\sigma$. Also, $\beta=1/T$, and we have used the notation $\Delta^-(/\Delta^+)=-i\gamma_5\epsilon_{f}\epsilon_{c}\Delta(/\Delta^*)$. Since the condensate of $u$ and $d$ quarks carries a net charge, there is a Meissner effect for ordinary magnetism, while photon-gluon mixing leads to a massless photon. In flavor$\otimes$color space in units of $\tilde{e}=\sqrt{3}ge/\sqrt{3g^2+e^2}$ the rotated charge matrix is given by $\tilde{Q}=Q\otimes I-I\otimes T^8/2\sqrt{3}$ ($T^3$ plays no role; the degeneracy of colors 1 and 2 ensures that there is no long-range gluon-3 field). In our case, this translates to $\tilde{Q}$ charges $u_{1,2}=1/2, d_{1,2}=-1/2, u_3=1, d_3=0$. With inert $s$-quarks, we also have $s_{1,2}=-1/2, s_3=0$. The gapped 2SC phase is $\tilde{Q}$-neutral, while overall charge neutrality of the matter requires a neutralizing background of strange quarks and/or electrons. In this article, we take the strange quark mass very large so that they do not play any dynamical role. We will limit ourselves to a discussion of the competition between condensates in a large magnetic field and not impose the charge neutrality and beta-equilibrium condition, which is known to stress the pairing and lead to gluon condensation and a strong gluo-magnetic field~\cite{Ferrer:2006ie}.

\vskip 0.2cm

Evaluation of the partition function and the thermodynamic potential $\Omega=-T{\rm ln}Z/V$ is facilitated by introducing 8-component Nambu-Gorkov spinors for each color and flavor of quark, leading to 

\beqy
&~&~{\rm ln}Z_{1,2}=\frac{1}{2}{\rm ln}\lbrace{\rm Det}(\beta G^{-1})\rbrace ;\quad  {\rm ln}Z_{3}=\frac{1}{2}{\rm ln}\lbrace{\rm Det}(\beta G_0^{-1})\rbrace\,;\\ \nonumber
G^{-1}&=&\bordermatrix{
&  & \cr
&[G_{0,\tilde{Q}}^+]^{-1} & \Delta^- \cr
&\Delta^+ & [G_{0,-\tilde{Q}}^-]^{-1} \cr}
\,,\quad G_0^{-1}=\bordermatrix{
&  & \cr
& [G_{0,\tilde{Q}}^+]^{-1} & 0\cr
&0 & [G_{0,-\tilde{Q}}^-]^{-1}\cr}\,,
\eeqy

where $[G_{0,\tilde{Q}}^{\pm}]^{-1}=(\hskip -0.1cm\not\!\partial\pm\mu\gamma^0+\tilde{e}\tilde{Q}\kern+0.30em /\kern-0.70em A-M)$. The determinant computation is simplified by re-expressing the $\tilde{Q}$-charges in terms of charge projectors in the color-flavor basis, following techniques applied for the CFL phase~\cite{Noronha:2007wg}. With the color-flavor structure unraveled, we can simplify the determinant computation further by introducing energy projectors~\cite{Huang:2001yw} and moving to momentum space using Fourier transformation, whereby we find

\beqy
{\rm ln}Z_{1,2}&=&{\rm Tr}_{c,f}\sum_{a}\sum_{p_0, {\bf p}}[{\rm ln}(\beta^2(p_0^2-(E_{\Delta,a}^+)^2)\beta^2(p_0^2-(E_{\Delta,a}^-)^2))]\,,\nonumber\\
{\rm ln}Z_3&=&{\rm Tr}_{f}\sum_{a}\sum_{p_0, {\bf p}}[{\rm ln}(\beta^2(p_0^2-(E_{p,a}^+)^2)\beta^2(p_0^2-(E_{p,a}^-)^2))]
\eeqy
\beqy
E_{\Delta,a}^{\pm}=\sqrt{(E_{p,a}^{\pm})^2+\Delta^2}\,,\quad E_{p,a}^{\pm}=E_{p,a}\pm\mu\,,\quad E_{p,a}=\sqrt{p_z^2+{\bf p}_{\perp,a}^2+M^2}\,,\,\,
 {\bf p}_{\perp,0}^2=p_x^2+p_y^2, \,\,{\rm else}\,\,{\bf p}_{\perp,\pm a}^2=2|a|\tilde{e}Bn\,.
\eeqy

The sum over $p_0=i\omega_k$ denotes the discrete sum over the Matsubara frequencies, $n$ labels the Landau levels in the magnetic field which is taken in the $\hat{z}$ direction and $a$=0,$\pm 1/2$ are the possible $\tilde{Q}$ charges of the quarks. 

\section{Gap Equations and Solution}
\label{sec_gap}

Using the following identity we can perform the discrete summation over the Matsubara frequencies
\begin{eqnarray}
\sum_{p_0}{\rm ln}[\beta^2(p_0^2-E^2)]=\beta[E+2T{\rm ln}(1+e^{-\beta E})\equiv\beta f(E)\,.
\end{eqnarray}

Then go over to the 3-momentum continuum using the following replacement

\begin{eqnarray}
\sum_{\bf p}\rightarrow{\rm V}\int\frac{d^3{\bf p}}{{(2\pi)}^3}~~{\rm where}~~{\textrm {V is the thermal volume of the system.}}
\end{eqnarray}

Finally, the zero-field thermodynamic potential can be expressed as

\begin{eqnarray}
\Omega_{B=0}=\frac{\sigma^{2}}{4G_{S}}+\frac{\Delta^{2}}{4G_{D}}-2\int_0^{\infty}\frac{d^3{\bf p}}{{(2\pi)}^3}[f(E_{p}^{+})+f(E_{p}^{-})+2f(E_{\Delta}^{+})+2f(E_{\Delta}^{+})]\,.
\end{eqnarray}

In presence of a quantizing magnetic field, discrete Landau levels suggest the following replacement

\begin{eqnarray}
\int_0^{\infty}\frac{d^3{\bf p}}{{(2\pi)}^3}\rightarrow\frac{|a|\tilde{e}B}{8\pi^{2}}\displaystyle\sum_{n=0}^{\infty}\alpha_{n}\int_{-\infty}^
{\infty}dp_{z}~~{\rm where}~~\alpha_{n}=2-\delta_{n0}\,,
\end{eqnarray}

where $\alpha_n$ is the degeneracy factor of the Landau levels (all levels are doubly degenerate except the zeroth level). The thermodynamic potential in presence of a magnetic field is given by

\begin{eqnarray}
\Omega_{B\neq 0}=\frac{\sigma^{2}}{4G_{S}}+\frac{\Delta^{2}}{4G_{D}}&-&\int_0^{\infty}\frac{d^3{\bf p}}{{(2\pi)}^3}[f(E_{p}^{+})+f(E_{p}^{-})]\nonumber\\
&-&\frac{\tilde{e}B}{8\pi^{2}}\displaystyle\sum_{n=0}^{\infty}\alpha_{n}\int_{-\infty}^
{\infty}dp_{z}[f(E_{p,1}^{+})+f(E_{p,1}^{-})+2f(E_{\Delta,\frac{1}{2}}^{+})+2f(E_{\Delta,\frac{1}{2}}^{-})]
\end{eqnarray}

In either case, we can now solve the gap equations obtained by minimizing the (zero-temperature) thermodynamic potential $\Omega$.

\begin{eqnarray}
{\textrm {Chiral gap equation}}:\frac{\partial\Omega}{\partial M}=0\,,\quad {\textrm {Diquark gap equation}}:
\frac{\partial\Omega}{\partial\Delta}=0
\end{eqnarray}

Since the above equations involve integrals that diverge in the ultra-violet region, we must regularize in order to obtain physically meaningful results. We choose to regulate these functions using a sharp cut-off (step function in $|{\bf p}|$), which is common in effective theories such as the NJL model~\cite{Schwarz:1999dj,Buballa:2001gj}, although one may also employ a smooth regulator~\cite{Alford:1998mk,Berges:1998rc, Noronha:2007wg} without changing the results qualitatively for fields that are not too large~\footnote{For example, a smooth cutoff was employed in~\cite{Noronha:2007wg} to demonstrate the De-Haas Van Alfen oscillations in the gap parameter at very large $B$.}. It should be noted that all free parameters of the system, viz., the bare quark mass, momentum cut-off and quark-antiquark coupling constant, are chosen at $B=0$ fitting the pion mass, pion decay constant and constituent quark mass in the vacuum. Our choice of cutoff implies
 
\begin{eqnarray}
\int_0^{\Lambda}\frac{d^3{\bf p}}{{(2\pi)}^3}\rightarrow\frac{|a|\tilde{e}B}{8\pi^{2}}\displaystyle\sum_{n=0}^{n_{max}}\alpha_{n}\int_{-\Lambda^{'}}^
{\Lambda^{'}}dp_{z}~~{\rm where}~~n_{max}={\rm Int}\Bigr[\frac{\Lambda^2}{2|a|\tilde{e}B}\Bigl];~~\Lambda^{'}=\sqrt{\Lambda^{2}-2|a|\tilde{e}Bn}\,.
\end{eqnarray}

From the fact that $p_{z}^{2}\geq 0$, we can calculate the maximum number of completely occupied Landau levels $n_{max}$. For magnetic field B$\lesssim 10^{18}$GeV$^2$, $n_{max}$ is of the order of 50 and the discrete summation over Landau levels becomes almost continuous. In that case, we recover the results of the zero magnetic field case, described in the next section. For fixed values of the free parameters, we were able to solve the chiral and diquark gap equations self-consistently, for $B=0$ as well as $B$ large. Before discussing our numerical results, we note the origin of the interdependence of the condensates. The chiral gap equation contains only $G_{S}$ which is determined by vacuum physics, but also depends indirectly on $G_D/G_S$ (a free parameter) through $\Delta$, which is itself dependent on the constituent $M=m_0+\sigma$. Our numerical results can be understood as a consequence of this coupling and the fact that a large magnetic field stresses the $\bar{q}q$ pair (same $\tilde{Q}$ charge, opposite spins implies anti-aligned magnetic moments) while strengthening the $qq$ pair (opposite $\tilde{Q}$ charge and opposite spins implies aligned magnetic moments). 

\section{Numerical analysis}
\label{nu_ana}

We first fix the NJL model's free parameters; the bare mass $m_u$=$m_d$=5.5 MeV, three-momentum cut-off $\Lambda$=$0.64$ GeV and $G_S$=$5.32~{\rm GeV}^{-2}$. These values are chosen to fit three vacuum quantities in the chiral limit: pion mass $m_{\pi}$=$134.98~{\rm MeV}$, pion decay constant $f_{\pi}$=$92.30~{\rm MeV}$ and the constituent quark mass $M(\mu=0)$=$330$ MeV~\cite{Klimt:1989pm,Huang:2001yw}. Although Fierz transforming one gluon exchange implies $R$=$G_D/G_S$=$0.75$ for $N_c$=3, the underlying interaction at moderate density is bound to be more complicated, therefore we choose to vary the strength of the diquark coupling channel to investigate the competition between the condensates as a function of the chemical potential. In the following we first discuss our results for the case with zero external field and then for non-zero external field.

\vskip 0.2cm

\begin{figure}
\includegraphics[width=80mm]{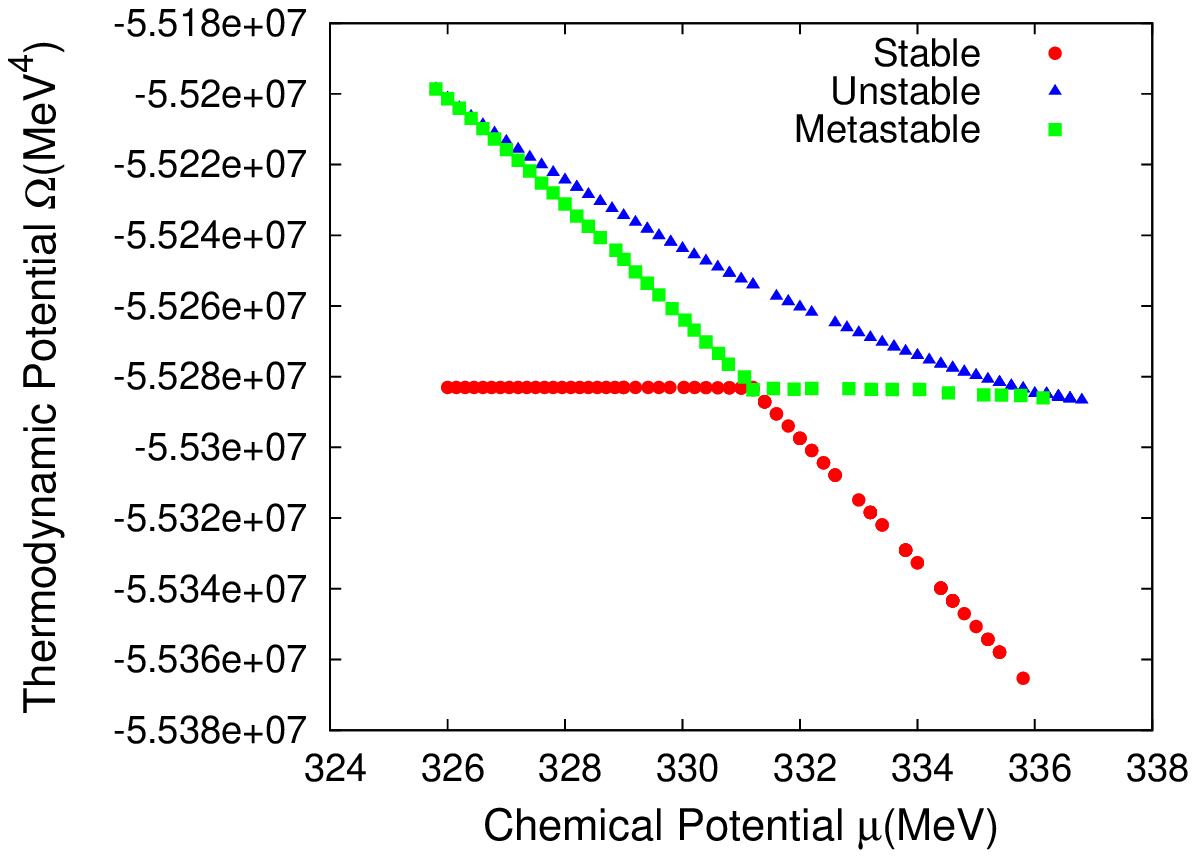}
\hspace{5mm} 
\includegraphics[width=80mm]{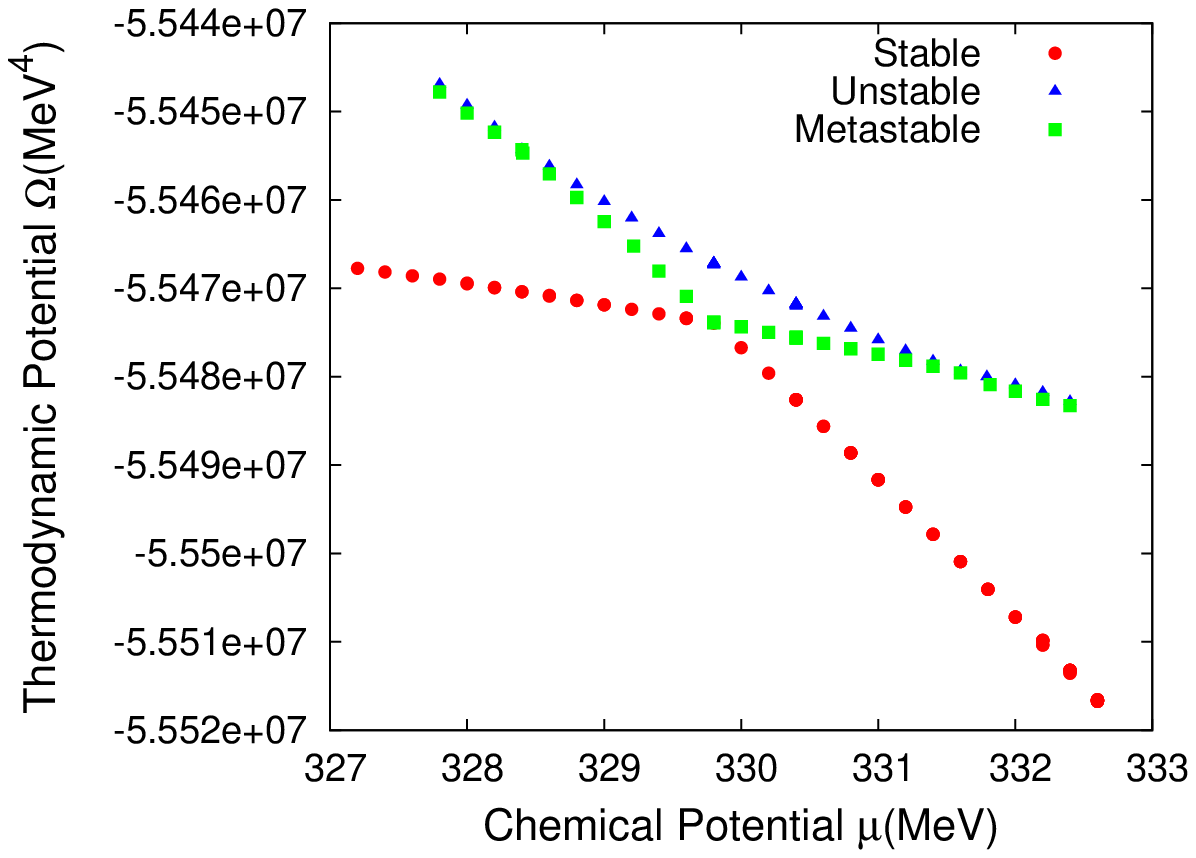}
\caption{\label{omega} Stable, metastable and unstable branches of $\Omega$ as determined for the case $B$=$0$ and $R$=$0.75$ (left panel) and $B$=$0.05$ GeV$^2$ and $R$=$0.75$ (right panel)}
\end{figure}

\begin{figure}
\caption{\label{gapfigs} Chiral gap (left panel) and Superconducting gap (right panel) as a function of shifted and scaled chemical potential $(1-\mu_c/\mu)$ for different values of $R$=$G_D/G_S$.}
\includegraphics[width=80mm]{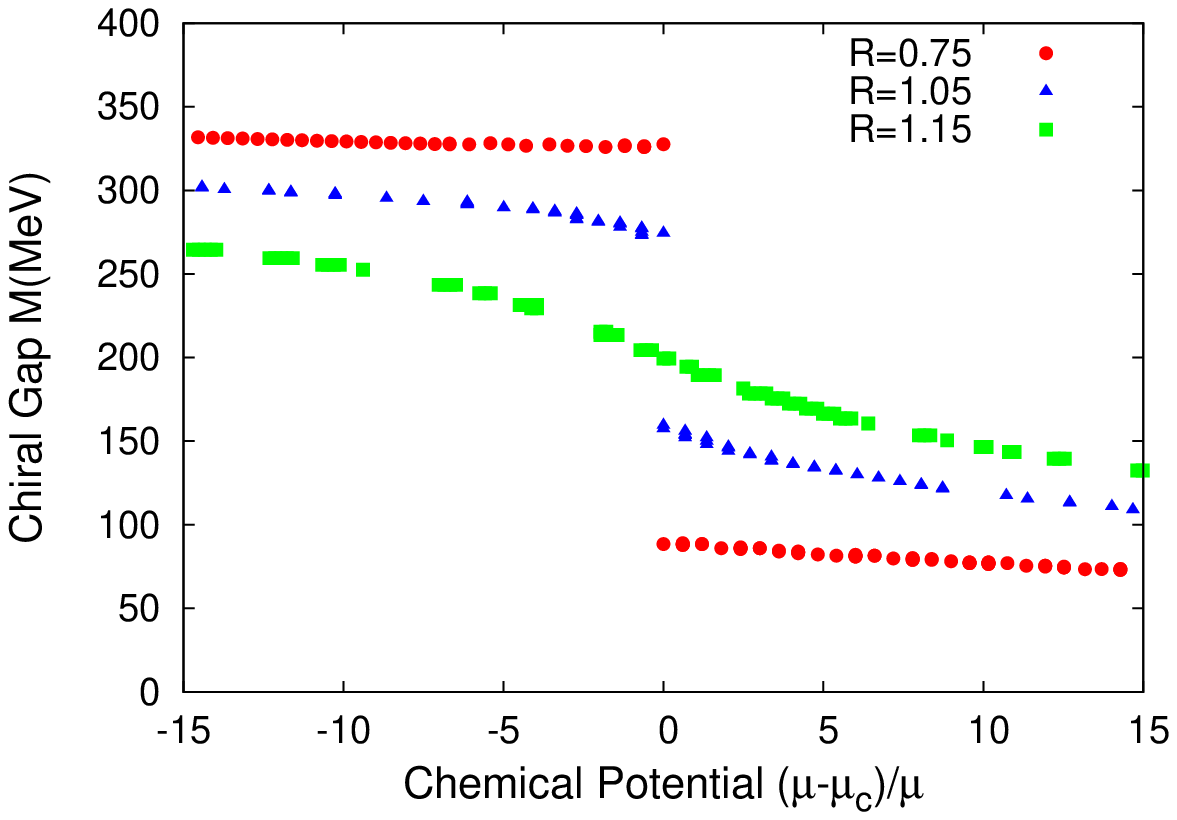}
\hspace{5mm} 
\includegraphics[width=80mm]{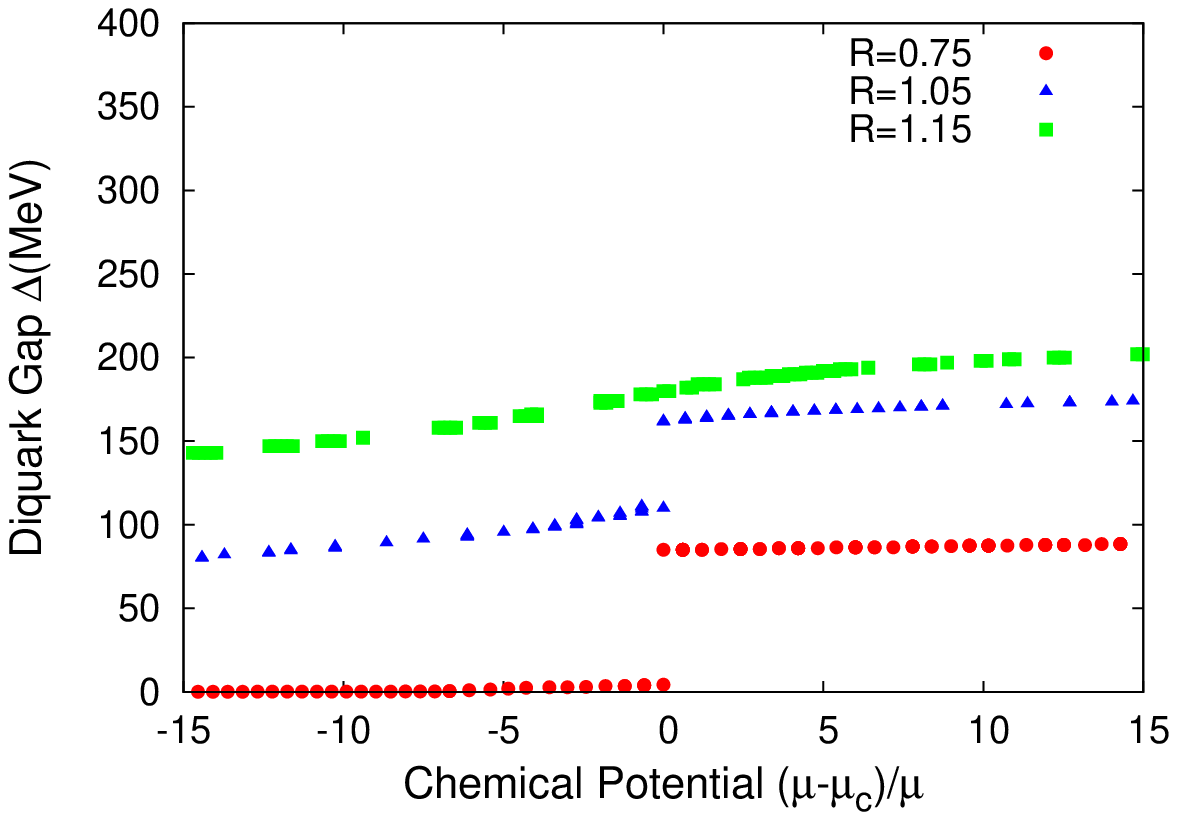}
\end{figure}

Case (a) {\bf No magnetic field}: 
As long as R is sufficiently small (we will quantify this below) and explicit 
breaking of chiral symmetry is slight, then for a narrow range of $\mu$ we get three solutions to the gap equations. These three solutions correspond to stable, metastable and unstable branches of the system. In Fig~\ref{omega} we plot the values of $\Omega$ corresponding to the three solutions obtained in the small $\mu$-window. The values of the gaps $(M,\Delta)$ for which $\Omega$ is the lowest correspond to the stable solution at any given density. The transition point or the critical chemical potential $\mu_c$ is the point where the first derivative of $\Omega$ (and the gaps) behave discontinuously. The superconducting gap abruptly becomes non-zero above $\mu_{c}$ while the chiral gap abruptly decreases implying partial restoration of chiral symmetry (since explicit breaking remains). The behavior of $\Omega$ and the gaps give us a clear indication of a first-order transition. In Fig~\ref{gapfigs} we plot the gaps versus $(\mu-\mu_c)/\mu$, which enables a convenient comparison, on the same graph, of the strength of the transition for different values of $R$=$G_D/G_S$. The actual value of this critical chemical potential, eg. for $R$=$0.75$ (which qualifies as sufficiently small) is $\mu_{c}$=$331.2$ MeV. Table~\ref{Rtable} displays $\mu_c$ and the magnitude of the jump in the order parameters at $\mu_c$ for different values of $R$.

\vskip 0.2cm

With increasing $R$, the superconducting gap appears at a smaller $\mu$ and rises smoothly from zero, until it becomes discontinuous at a critical chemical potential where the chiral gap changes discontinuously (albeit with a smaller jump than before). The broken lines in Fig.~\ref{gapfigs}, for $R=1.05$, clearly demonstrate this coupled feature of the two condensates, with a first-order chiral transition affecting the superconducting gap (see also Huang et al.~\cite{Huang:2001yw} for this feature). While not obvious from Fig.~\ref{gapfigs} due to the shift $\mu\rightarrow(\mu-\mu_c)/\mu$, the chiral phase transition occurs at a lower $\mu$ than before. This picture does not change qualitatively until we go above a critical value $R$=$R_c$=1.11; i.e, while $R<R_c$, the discontinuities in $m$ and $\Delta$ remain non-zero but decrease smoothly. At $R=R_c$ the metastable and unstable regions vanish completely. This qualifies it to be a second order phase transition. Huang et al.~\cite{Huang:2001yw} have termed the region where the condensates co-exist and vary continuously, as the mixed broken phase, since both chiral and (global) color symmetry are broken here. While it should not be confused with a genuine mixed phase, since the free energy admits a unique solution to the gap equations in this regime, the width of this overlap region increases with increasing $R$. Above $R=R_c=1.11$, $\sigma$ and $\Delta$ are smoothly varying resulting in a smooth crossover. However there is always a pseudo-transition point ($\mu_{cp}$) around which fluctuations/variations of both the condensates are sharply peaked. The width of these peaks broaden with further increase of $R$ and $\mu_{cp}$ moves towards the left with increasing $R$. These results for $B=0$ agree qualitatively with the results of Huang et al.~\cite{Huang:2001yw} with minor quantitative differences in $\mu_c,\mu_{cp}$ at less than $1.5\%$ level.

\begin{figure}
\caption{\label{bgapfigs} Chiral gap (left panel) and Superconducting gap (right panel) as a function of chemical potential $\mu$ for $B$=0 and $B\sim 10^{18}$G. The value of $R$=$G_D/G_S$=1.2 in both cases.}
\includegraphics[width=80mm]{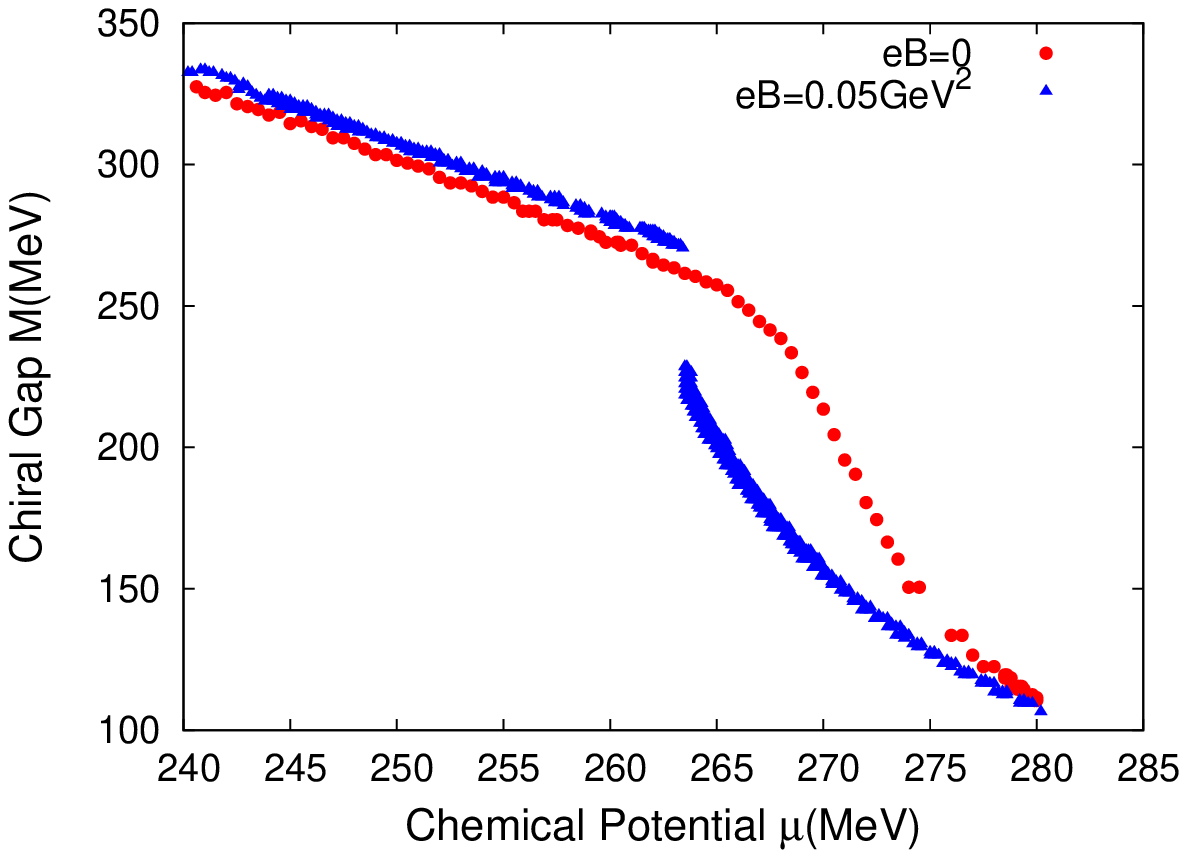}
\hspace{5mm} 
\includegraphics[width=80mm]{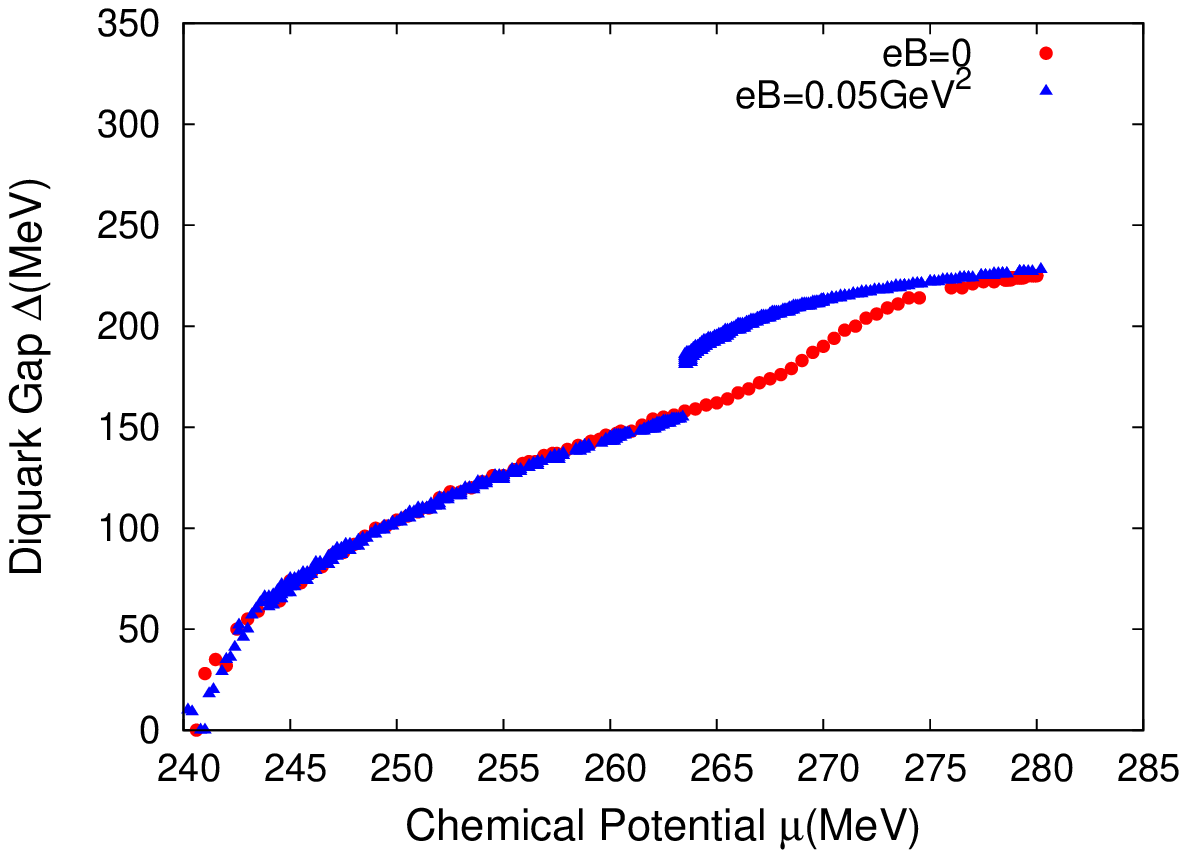}
\end{figure}

\vskip 0.2cm

\begin{center}
\begin{table}
\caption{\label{Rtable} Critical chemical potential $\mu_c$, jumps in the chiral ($\delta_{\sigma}$) and superconducting ($\delta_{\Delta}$) order parameters at $\mu_c$ for zero and large $B$-field for various values of $R$=$G_D/G_S$. The nature of the transition is also indicated.}

\vskip 0.2cm

\begin{tabular}{|p {1cm}|p {1.5cm}|p {1.5cm}|p {1.5cm}|p {2cm}|p {1.5cm}|p {1.5cm}|p {1.5cm}|p {2cm}|}
\hline
R&\multicolumn{4}{|l|}{~~~~~~~~~~~~~~~~~~~~~~~~~~eB=0}&\multicolumn{4}{|l|}{~~~~~~~~~~~~~~~~~~~~~~~~eB=0.05Ge${\rm V}^2$}\\
\cline{2-9}
 &$\delta_{\sigma}$(MeV)&$\delta_{\Delta}$(MeV)&$\mu_{c}$(MeV)&Nature&$\delta_{\sigma}$(MeV)&
 $\delta_{\Delta}$(MeV)&$\mu_{c}$(MeV)&Nature\\
\hline
0.75&237.9&80.6&331.2&First order&158.4&53.4&329.5&First order\\
\hline
1.05&115.1&51.2&295.3&First order&105.6&66.2&291.5&First order\\
\hline
1.11&0&0&284.9&Second order&67.1&44.4&280.7&First order\\
\hline
1.15&Smooth&Smooth&279.4&Crossover&57.0&32.8&273.3&First order\\
\hline
1.20&Smooth&Smooth&269.7&Crossover&40.7&25.5&263.2&First order\\
\hline
\end{tabular}
\end{table}
\end{center}

\vskip -0.75cm

Case (b) {\bf Large Magnetic field}: Our self-consistent method of solving the gap equations is now applied to study the competition between the condensates in presence of a strong magnetic field, at maximum of the order of $10^{18}$GeV$^2$~\footnote{We have ignored magnetic catalysis of chiral symmetry breaking, important for extremely large fields $B\sim 10^{20}$G~\cite{Gusynin:1995gt}}. Since the chiral condensate involves quark spinors of opposite spin and same $\tilde{Q}$-charge, it is stressed by large magnetic fields. The diquark condensate, with opposite spin and $\tilde{Q}$-charge, is strengthened. Thus, we expect and indeed observe, as seen from Fig.~\ref{bgapfigs}, a strengthening of the competition between the two condensates, resulting in a qualitative change from the $B=0$ case. For moderate values of $R\lesssim 1.0$, the sharp first order jumps in the $B$=0 case are weakened and the metastable region shrinks. The effect is more dramatic in the crossover region (large $R$), where the smooth behavior is replaced by a first-order transition (see Fig~\ref{bgapfigs}). With our choice $R=1.2$ (employed in Fig.~\ref{bgapfigs}), we find that a smooth crossover in the $B=0$ case at $\mu_{\rm cp}\sim 270$ MeV gives way to a discontinuity in the chiral and superconducting gaps at $\mu_c\sim 263$ MeV for $eB=0.05$ GeV$^{2}$ ($B=8.5\times 10^{18}$G). There is only a very small change in the transition density for superconductivity and (partial) chiral restoration (see Table~\ref{Rtable}). However, the mixed broken phase of Huang et al.~\cite{Huang:2001yw} is disrupted. The simultaneous appearance of the discontinuity in the chiral and superconducting gap for large magnetic field case, at almost the same $\mu$ where the condensates have their most rapid variation in the $B$=$0$ case, is a physical feature and is also cutoff insensitive. Finally, we have checked that magnetic fields $eB\lesssim 0.01$~GeV$^2$ ($B\lesssim 10^{18}$G) do not notably alter the competition between the condensates from the zero magnetic field case.

\section{Conclusions}
\label{sec_conc}

We have studied the effect of a large magnetic field on the chiral and
diquark condensate in a two-flavor superconductor using the NJL model.
We have implemented a self-consistent scheme to determine the condensates,
by numerically iterating the coupled (integral) equations for the chiral
and superconducting gap. We have obtained results for the nature of the competition between these condensates in two cases: (a) at zero magnetic field, where
our results are qualitatively the same as those of Ref.~\cite{Huang:2001yw},
and small quantitative differences arise most likely due to the choice of free parameters and a different numerical scheme; and (b) at large magnetic fields, where we find that the competition increases strongly, causing a discontinuity in
the gaps, and disrupting the ``mixed broken phase''. This is a result of the 
modified free energy of the quarks in the condensate when subjected to a 
magnetic field. For magnetic fields as large as $B\sim 10^{18}$G, the 
anti-aligned magnetic moments of the quarks in the chiral
condensate change the smooth crossover of the chiral transition to a sharp 
first order transition. The diquark gap also becomes discontinuous at this 
point. For magnetic fields $B\leq 10^{18}$G, there is no significant effect of the magnetic field on the competition between the condensates and zero-field results apply.

\vskip 0.2cm
 
Although we have used an effective model with a sharp cutoff and 
omitted certain important considerations like the finite mass of the 
strange quark and neutrality constraints on the matter which can induce additional condensates that enter the competition, it is interesting to speculate on possible physical consequences of the transition induced by the large magnetic field, since it is a strong and probably general feature of competing condensates. At some fixed large value of the local field $B$ and in the small density
window of the metastable region, cooling of two-flavor quark matter below the 
superconducting $T_c$ can result in the formation of domains or nuggets of superconducting regions with different values for the gap, even at constant pressure (a detailed analysis requires assessment of surface and screening effects in superconducting quark matter~\cite{Jaikumar:2005ne,Alford:2006bx}). The chiral gap (or constituent mass) in this mixed phase shows the same feature, hearkening back to the DCC idea~\cite{Rajagopal:1993ah}. If the system is formed in a metastable state and then drops to the true minimum of the free energy, nucleation of chirally restored and superconducting droplets can happen simultaneously. This is a heterogeneous nucleation process since superconducting domains can already exist. We can also imagine that external magnetic fields of order $eB/\mu^2\sim 1$ show some local variation on the microscopic scale in the initial stages of the formation of the dense neutron star. In this case, magnetic domains with different magnetization can form. Such kinds of nucleation and domain formation will release latent heat that might be very large owing to the large value of the magnetic field, serving as a large internal engine for possible energetic events on the surface of the neutron star. Such internal mechanisms are unlikely to occur in a pure neutron star without a quark core~\cite{Broderick:2001qw} and could have important applications for high-energy astrophysics~\cite{Ouyed:2005dz}. In addition, it has been recently shown that large magnetic fields in strange quark matter introduce anisotropies in bulk viscosities, which can change the stability region of the r-mode in such stars~\cite{Huang:2009ue}. A similar effect may arise in magnetized superconducting two-flavor or 2SC+s quark matter, another important link between the microscopic and astrophysical consequences of a large magnetic field in quark matter in the core of neutron stars.

\vskip 0.2cm

\section*{Acknowledgments}

We are grateful to Neeraj Kumar Kamal for assistance with certain numerical 
aspects of this work.

\end{document}